\documentclass[prl,twocolumn,floatfix,superscriptaddress]{revtex4}
\usepackage{times,amsmath,amsfonts,amssymb,latexsym}
\usepackage{graphicx,epsf}

\newcommand{\ket}[1]{| #1 \rangle}

\newcommand{\be}{\begin{equation}}
\newcommand{\ee}{\end{equation}}
\newcommand{\ba}{\begin{eqnarray}}
\newcommand{\ea}{\end{eqnarray}}

\newcommand{\ignore}[1]{}
\def\CC{{\rm\kern.24em \vrule width.04em height1.46ex depth-.07ex
    \kern-.30em C}}
\def\P{{\rm I\kern-.25em P}}
\def\RR{{\rm
         \vrule width.04em height1.58ex depth-.0ex
         \kern-.04em R}}

\def\bbbc{{\mathchoice {\setbox0=\hbox{$\displaystyle\rm C$}\hbox{\hbox
to0pt{\kern0.4\wd0\vrule height0.9\ht0\hss}\box0}}
{\setbox0=\hbox{$\textstyle\rm C$}\hbox{\hbox
to0pt{\kern0.4\wd0\vrule height0.9\ht0\hss}\box0}}
{\setbox0=\hbox{$\scriptstyle\rm C$}\hbox{\hbox
to0pt{\kern0.4\wd0\vrule height0.9\ht0\hss}\box0}}
{\setbox0=\hbox{$\scriptscriptstyle\rm C$}\hbox{\hbox
to0pt{\kern0.4\wd0\vrule height0.9\ht0\hss}\box0}}}}
\def\bbbz{{\mathchoice {\hbox{$\sf\textstyle Z\kern-0.4em Z$}}
{\hbox{$\sf\textstyle Z\kern-0.4em Z$}}
{\hbox{$\sf\scriptstyle Z\kern-0.3em Z$}}
{\hbox{$\sf\scriptscriptstyle Z\kern-0.2em Z$}}}}

\begin{document}
\title{Hidden entanglement at the Planck scale: loss of unitarity and the information paradox}
\author{Michele Arzano}
\email{marzano@perimeterinstitute.ca}
\affiliation{Perimeter Institute for Theoretical Physics\\
31 Caroline St. N, N2L 2Y5, Waterloo ON, Canada}
\author{Alioscia Hamma}
\email{ahamma@perimeterinstitute.ca}
\affiliation{Perimeter Institute for Theoretical Physics\\
31 Caroline St. N, N2L 2Y5, Waterloo ON, Canada}
\affiliation{Massachusetts
Institute of Technology, Research Laboratory of Electronics\\
77 Massachusetts Ave. Cambridge MA 02139}
\author{Simone Severini}
\email{simoseve@gmail.com}
\affiliation{Institute for Quantum Computing and Department of Combinatorics \& Optimization\\
University of Waterloo, 200 University Av. W, N2L 3G1, Waterloo ON,
Canada}

\begin{abstract}
We discuss how relaxing the requirement of locality for quantum fields can equip the Hilbert space of the theory
with a richer structure in its multi-particle sector. A physical consequence is the emergence of a ``planckian" mode-entanglement, invisible to an observer that cannot probe the Planck scale. To the same observer, certain unitary processes would appear non-unitary.  We show how entanglement transfer to the additional degrees of freedom can provide a potential way out of the black hole information paradox.
\end{abstract}

\maketitle

{\em Introduction.---} The phenomenon of black hole quantum radiance
discovered by Hawking \cite{Hawking:1974sw} is a typical effect due to the propagation of quantum fields on curved space.  Vacuum fluctuations on a Schwarzschild background lead to the creation of entangled pairs of particles inside and outside the horizon. The spectrum of particles observed outside the horizon is the {\it Hawking radiation}. As it is well known, the semiclassical understanding of black hole evaporation seems to be in contrast with the rules of ordinary quantum mechanics \cite{Hawking:1976ra}. An evaporating black hole loses its mass, shrinks and
eventually disappears. The emitted radiation is in a thermal state, {\it i.e.}, a highly mixed state. If the black hole was formed starting from a pure quantum state, at the end of the evaporation process, all we are left
with is a mixed state. This conclusion violates the unitarity of
evolution of closed quantum states, one of the basic pillars of quantum
mechanics. Such a contradiction is known as the {\em black hole
information paradox}. The problem can be clearly stated in the
framework of quantum information theory. The radiation consists of entangled pairs, that can be
approximatively seen as Bell pairs. One member of the pair is
inside the horizon, while the other one is outside. All these qubits are entangled with respect to the
bipartition ({\em inside the horizon})-({\em outside the horizon}). When the evaporation process is completed, the particles originally outside the horizon would be {\it entangled with nothing} because the bipartition has disappeared. It follows that either the fundamental theory is not
unitary, or the entanglement has to decrease to zero during the evaporation.

Proposals to solve the above problem have invoked a leak of information through the horizon via some non-local effect (see, {\it e.g.} \cite{Giddings:2006sj}). Indeed, locality appears to be only an
approximate concept in a quantum gravity framework. This is mainly
due to the difficulty in constructing observables that, in a certain
limit, reproduce the familiar observables of local quantum field theory (QFT)
\cite{Rovelli:1990ph, Giddings:2005id, Dittrich:2005kc}. Moreover, holographic arguments suggest that a strict locality
requirement for quantum observables can not survive the introduction
of dynamical gravitational effects \cite{ArkaniHamed:2007ky}.

In this letter, we show how violations of
locality in QFT can endow the Hilbert space of the
theory with a richer structure and with degrees of freedom accessible only to processes which
can resolve planckian scales. The entanglement present during the black hole evaporation process can then be transferred from the original {\em in-out} bipartition
to a new bipartition outside the horizon. We argue that mode entanglement between ``planckian" modes and
ordinary ``macroscopic" modes could help escaping the puzzling
conclusions of the information paradox.

{\em Non-locality, quantum symmetries, and momentum dependent
statistics.---} If one is willing to adhere to some non-locality
paradigm, as discussed above, then the natural question is whether there exist some
modification of local QFT, whose features are
qualitatively different in contexts like, for example, particle
production in the presence of an horizon. Lower dimensional QFT models provide some useful insights in this regard.
Non-local currents are well studied objects in 2-dimensional
QFT and their associated charges exhibit a non-trivial algebraic
structure which is described in terms of quantum groups \cite{Bernard:1990ys}.  The
emergence of algebraic structures more complex than ordinary Lie
algebras can be traced back to the possibility of relaxing one of
the assumptions of the Coleman-Mandula theorem, namely the
requirement that the charges (symmetry generators) act on
multi-particle states as derivatives, {\it i.e.}, following a generalized
Leibnitz rule.  As it will be discussed in what follows, such
property is intimately related to the Lie algebra structure of the
generators of symmetries, their (adjoint) action on observables of
the theory, and ultimately, to locality. It has been argued \cite{Arzano:2007nx} that
relaxing the requirement of locality in a controlled way naturally
leads to generalizing the description of the space-time symmetries
of quantum fields to the realm of quantum groups.  Quantum fields
with quantum group symmetries provide in turn a possible framework
for non-commutative field theories (see, {\it e.g.} \cite{AmelinoCamelia:2001fd, Chaichian:2004za}). Here we elaborate on the argument given in \cite{Arzano:2007nx} and, focusing on a specific example, we
discuss the non-trivial features that distinguish the new model from
ordinary local QFT.

The Hilbert space of a linear scalar field is usually described in
terms of ``plane-wave" state vectors (the ``wave-modes" of
the field) labeled by their eigenvalues
with respect to the space-translation generators.  In the usual picture, such translation generators are
identified with the charges associated to the translational symmetry
of the classical theory.  In particular such charges are suitable
integrals of the quantized Noether currents. The fact that they
generate the symmetry is most easily seen using the Ward identity involving the current and a product of $n$-fields \cite{DiFrancesco:1997nk}.
Integrating such identity, one finds that the action of the
translation generators $P_{\mu}$ on a basis ``mode" vector
$|\vec{k}\rangle=a^{\dagger}(\vec{k})\,|0\rangle$ is given in terms
of the conserved charges $\Pi_{\mu}$ by
%, in terms of creation operators the action on the one-particle state, by
\begin{equation}\label{pm}
P_{\mu}\vartriangleright|\vec{k}\rangle=k_{\mu}|\vec{k}\rangle=[\Pi_{\mu},
a^{\dagger}(\vec{k})]\,|0\rangle \, .
\end{equation}
From a mathematical point of view, the commmutator on the right hand side
represents the adjoint action of the (Lie) algebra of translation
generators on the algebra of creation operators.  Additivity of the translation generators on
multiparticle states immediately follows from the well known
property of the commutator
\begin{equation}
[A, B_1B_2...B_n]=\sum_{i=1}^n B_1...[A,B_i]...B_n\, .
\end{equation}
In 2d quantum integrable models, the complete non-perturbative
knowledge of the S-matrix leads to the construction of non-local
charges whose action on multi-particle states is non-additive \cite{Bernard:1990ys}.
Such behavior reveals an algebraic structure described by quantum
algebras instead of ordinary Lie algebras.  In \cite{Arzano:2007nx} it was argued that
an analogous situation might occur for the generators of spacetime
symmetries in QFT when their associated charges
experience an intrinsic non-locality determined by dynamical
gravity. To understand how quantum deformations of
the algebra provide a generalization of the Leibnitz rule and of the
adjoint action (\ref{pm}) we note that for a translation generator the
action on a tensor product (two-particle) state is described by
the operator $\Delta^{(0)}(P_{\mu})=P_{\mu}\otimes1+1\otimes
P_{\mu}$. In general, given a quantum deformed algebra, with
deformation parameter $h$, such operator becomes
$\Delta(P_{\mu})=\Delta^{(0)}(P_{\mu})+h\Delta^{(1)}(P_{\mu})+O(h^2)$
with the important property that given the ``flip" operator $\sigma:
\sigma(a\otimes b)=b\otimes a$ one has $\Delta(P_{\mu})\circ\sigma\neq
\Delta(P_{\mu})$.   The adjoint action is defined by
\begin{equation}
\label{adj2} P_{\mu} \vartriangleright |\vec{k}\rangle=(id\otimes
S)\Delta(P_{\mu})\diamond a^{\dagger}(\vec{k})|0\rangle \, ,
\end{equation}
where $(F\otimes G) \diamond a\equiv FaG$ and
$S(P_{\mu})=S^{(0)}(P_{\mu})+h\,S^{(1)}(P_{\mu})+O(h^2)$ with
$S^{(0)}(P_{\mu})=-P_{\mu}$.  Note how $S(P_{\mu})$ provides a deformed inversion
map (involution).  In the case $h=0$, one recovers the usual adjoint
action (\ref{pm}) of the Poincar\'e (Lie) algebra. The deformed
quantum algebra captures the gravity-induced non-local effects that
would manifest in Planck-scale suppressed corrections to the
classical adjoint action (\ref{pm})
\begin{eqnarray}
\langle 0\,|P_{\mu }\vartriangleright |\vec{k}\rangle  &\equiv &\langle
0\,|[P_{\mu },a^{\dagger }(\vec{k})]|\,0\rangle   \label{nladj} \\
&&+\alpha _{1}E_{p}^{-1}F^{(1)}(\vec{k})+O(E_{p}^{-2})  \nonumber
\end{eqnarray}
if one identifies the deformation parameter $h$ with $E_p^{-1}$, the inverse Planck energy \cite{Arzano:2007nx}.\\
In this letter we will focus on the particular example of the
$\kappa$-Poincar\'e algebra, a well known  quantum Poincar\'e
algebra with deformation parameter $h=1/\kappa$ \cite{Majid:1994cy}.  For spatial momenta one has
\begin{equation}
\label{copro} \Delta(P_i)=P_i\otimes 1+ e^{-P_0/\kappa}\otimes P_i ,
\end{equation}
while the time-translation generator still acts according to the
Leibnitz rule.  Only recently a satisfactory understanding of the
basic properties of free scalar quantum fields with such
$\kappa$-symmetries has been reached \cite{Arzano:2007ef}.  The most
peculiar feature of the Hilbert space of
$\kappa$-bosons is that they obey a momentum dependent statistics
arising from the non-trivial structure of the multi-particle sector
of the theory. Such structure is a direct consequence of the
``non-symmetric" form of (\ref{copro}). Take for example the two
states $|\vec{p}_1\rangle\otimes|\vec{p}_2\rangle$ and
$|\vec{p}_2\rangle\otimes|\vec{p}_1\rangle$. Unlike the undeformed
case they now have two {\it different} eigenvalues of the linear
momentum, respectively $p_{i1}+ e^{-p_{01}/\kappa} p_{i2}$
and $p_{i2}+ e^{-p_{02}/\kappa} p_{i1}$. Clearly the usual
``symmetrized" two-particle state
\begin{equation}
1/\sqrt{2}(|\vec{p}_1\rangle\otimes
|\vec{p}_2\rangle+|\vec{p}_2\rangle\otimes |\vec{p}_1\rangle)
\end{equation}
is no longer an eigenstate of the momentum operator.  Rather, given
two modes $\vec{p}_1$ and $\vec{p}_2$, we have two different
$\kappa$-symmetrized two-particle states
\begin{eqnarray}
|p_{1}p_{2}\rangle _{\kappa } &=&\frac{1}{\sqrt{2}}\left[ |\vec{p}%
_{1}\rangle \otimes |\vec{p}_{2}\rangle +|(1-\epsilon _{1})\vec{p}%
_{2}\rangle \right.  \\
&&\otimes \left. \,|(1-\epsilon _{1}(1-\epsilon _{2}))^{-1}\vec{p}%
_{1}\rangle \right]   \nonumber
\end{eqnarray}%
\begin{eqnarray}
|p_{2}p_{1}\rangle _{\kappa } &=&\frac{1}{\sqrt{2}}\left[ |\vec{p}%
_{2}\rangle \otimes |\vec{p}_{1}\rangle +|(1-\epsilon _{2})\vec{p}%
_{1}\rangle \right.  \\
&&\otimes \left. \,|(1-\epsilon _{2}(1-\epsilon _{1}))^{-1}\vec{p}%
_{2}\rangle \right]   \nonumber ,
\end{eqnarray}
with $\epsilon_i=\frac{|\vec{p}_i|}{\kappa}$ \cite{Arzano:2007ef}.\\
{\em The fine structure of $\kappa$-Hilbert space and mode
entanglement.---}
The construction of QFT using the deformed symmetries described above induces a radical modification in the number of degrees of freedom of the system.  We show here how the new fine structure of the Hilbert space is determined by the splitting in the linear momenta between quantum states with different mode ordering.  For the sake of simplicity, let us focus on the two-mode state case.
We have already seen that the states $|p_1p_2\rangle_{\kappa}$ and
$|p_2p_1\rangle_{\kappa}$ are not identical. In fact they are
orthogonal for every finite $\kappa$, as it can be verified by taking the scalar product
\begin{eqnarray}
\nonumber \langle p_1p_2\,|\,p_2p_1\rangle_{\kappa}\simeq\frac{1}{2}
\delta^{(3)}\left(\epsilon_2\vec{p}_1\right)\times \\
\delta^{(3)}\left(\frac{\epsilon_1(1-\epsilon_2)}{1-\epsilon_1(1-\epsilon_2)}\vec{p}_2\right)+
1\leftrightarrow 2\ .
\end{eqnarray}
In the undeformed case ($\kappa \rightarrow \infty$)
we recover the indistinguishability of the two states, {\it i.e.},
$\langle p_2p_1|p_1p_2 \rangle_{\kappa} =1$. The two states can be
distinguished by measuring the splitting in their linear momenta.
The linear momentum for the state $\ket{p_ip_j}_{\kappa}$ is
$\vec{P}_{ij}=\vec{p}_i+e^{-p^0_i/\kappa}\vec{p}_j$, where $p_i^0
=-\kappa\log(1-|\vec{p}_i|/\kappa)$. The resulting splitting is thus \be
|\Delta\vec{P}_{12}|\equiv |\vec{P}_{12}-\vec{P}_{21}| =\frac{1}{\kappa}
|\vec{p}_1|\vec{p}_2|-\vec{p}_2|\vec{p}_1||\le\frac{2}{\kappa}|\vec{p}_1||\vec{p}_2|.\ee
Since such splitting is a quantity of the order
$|\vec{p}_i|^2/\kappa$, we see that the states become indeed
indistinguishable in the limit of $\kappa\rightarrow\infty$. In
practice, this happens when the resolution of our measurements is
much inferior than the splitting. We can see that only for very high
momenta the splitting becomes relevant for an observer that has a
reasonable resolution. Nevertheless, the Hilbert space has now a new
degree of freedom: the mode of being of the type $|p_1p_2\rangle
_{\kappa}$ or $|p_2p_1\rangle _{\kappa}$. Formally, the Hilbert
space acquires the following tensor product structure: $\mathcal
H^2_{\kappa} \cong \mathcal S_2\mathcal H^2 \otimes \mathbb{C}^2$,
where $\mathcal S_2 \mathcal H^2$ is the standard symmetrized 2-particle Hilbert space. We
can write the two states as
\begin{eqnarray}
|E\rangle\otimes|0\rangle &=& |p_1p_2\rangle _{\kappa}\\
|E\rangle\otimes|1\rangle &=&|p_2p_1\rangle _{\kappa}
\end{eqnarray}
where $E=E(\vec{p}_1)+E(\vec{p}_2)$. The $\mathbb{C}^2$ degrees of
freedom can only be seen under a planckian ``magnifying glass". Far
from the Planck scale, this variable is hidden. Many interesting
phenomena can occur  because of these additional degrees of freedom.
First, one can have mode entanglement. The state of the
superposition of two total ``classical" energies
$E_A=E(\vec{p}_A)+E(\vec{q}_A)$ and $E_B=E(\vec{p}_B)+E(\vec{q}_B)$
can be entangled with the additional hidden modes. We could have,
for instance, a superposition of the form $|\Psi\rangle =
1/\sqrt{2}(|E_A\rangle\otimes|0\rangle
+|E_B\rangle\otimes|1\rangle)$.

In the case of $n$ particles with distinct momenta we will have $n!$ orthogonal plane wave
states corresponding to the $n!$ different orderings. In general, given $n$-particles, of which $n_i$ have momentum $p_i$ with $\sum_{i=1}^R n_i=n$, we denote their $\kappa$-symmetrized Hilbert space with $\mathcal H^{(\vec{n}_R)}_\kappa$ where $\vec{n}_R \equiv (n_1,...,n_R)$. The number of orthogonal states in $\mathcal H^{(\vec{n}_R)}_\kappa$ is $N(\vec{n}_R)=n!/\prod_{i=1}^R n_i!$.
For such space we can write the following decomposition
\be\label{tensprod}
\mathcal H^{(\vec{n}_R)}_\kappa
\cong \mathcal S_n\mathcal H^n \otimes \mathbb{C}^{N(\vec{n}_R)}
\ee
where $\mathcal{H}^n=\underbrace{\mathcal{H}^{1}\otimes\mathcal{H}^{1}...\otimes\mathcal{H}^{1}}_\text{n-times}$ and $\mathcal S_n$ sums over all permutations of $n$ objects. Here, $\mathcal{H}^{1}$ is the usual one-particle Hilbert space.
The $n$-particle Hilbert space will be given by $\mathcal H^{n}_{\kappa} = \oplus_{n_1,...,n_R}\mathcal H^{(\vec{n}_R)}_\kappa$ and we can thus have states containing up to $\log_2 N$ ebits with respect to the tensor product structure in (\ref{tensprod}).

{\em Loss of unitarity and the black hole information
paradox.---} What are the implications of this construction for an observer that is unable to resolve the degrees
of freedom in the right hand side of the tensor product in Eq.(\ref{tensprod})? To answer this question, consider a quantum system evolving unitarily,
through a Hamiltonian $H$ defined on the Hilbert space $\mathcal
H^n_{\kappa}$. The quantum evolution of the
system, described by the density matrix $\rho(t)$, would read \be
\rho(t) = U(t) \rho(0) U^\dagger(t) \ee where $U(t) = \mathcal T
e^{-iHt}$. Now suppose we start with a pure state $\rho(0)$ factorized with respect to the bipartition in $\mathcal H^{(\vec{n}_R)}_{\kappa}$. If the unitary
$U(t)$ acts as an entangling gate, the state $\rho(t)$ will be entangled.
In order for this to happen, it is necessary that the Hamiltonian addresses
simultaneously the degrees of freedom in the bipartition of $\mathcal H^{(\vec{n}_R)}_{\kappa}$.
If the observer is not
able to resolve the planckian degrees of freedom, what she will see
is the reduced system obtained by tracing out the degrees of freedom
in $\mathbb{C}^{N(\vec{n}_R)}$. Notice that if the pure state $\rho(0)$ is
separable, also the reduced system
$\rho_{obs}(0)=\mbox{Tr}_{Pl}\rho(0)$ will be pure. As the system
evolves, we have \be \rho_{obs}(t) = \mbox{Tr}_{Pl}\rho(t)=
\mbox{Tr}_{Pl}\left[ U(t) \rho(0) U^\dagger(t) \right].\ee
For the observer, this evolution is not unitary. We have in fact started
with a pure state $\rho_{obs}(0)$ and ended up with a mixed
state $\rho_{obs}(t)$.
This is analogous to evolution in open quantum systems. The reduced system
evolves according to a completely positive map, taking density
matrices into density matrices, but the process is not unitary. The
entanglement with the environment makes the system dissipative \footnote{A somewhat similar framework has been proposed in L.~J.~Garay, Phys.\ Rev.\ Lett.\  {\bf 80}, 2508 (1998) [arXiv:gr-qc/9801024], where it was argued that spacetime foam effects, modeled as non-local interactions, would cause a loss of coherence in quantum evolution.}. Here, the environment is simply constituted by those internal
degrees of freedom that the observer cannot measure. The environment is not ``out'' there, but
``in'' there. What we have obtained is an apparent loss of unitarity, related to the
incapability of the observer to resolve the fine structure of the deformed Hilbert space of the theory.  We could prepare a pure state, send it into a unitary channel, which is able to affect the planckian degrees of freedom, and measure the purity of the final state. If the final state is not pure, it is because it is mixed with
the degrees of freedom which we cannot resolve. But what could be such a channel?

One possibility is to consider a black hole. If we believe in quantum mechanics, a black hole that evaporates is a unitary channel, but nevertheless transforms a pure state into a mixed state. This is the black hole information paradox.  The essence of the paradox can be
seen in the nature of the emitted radiation consisting of
entangled pairs (for a recent review see, {\it e.g.}, \cite{mathur}).
The mechanism for Hawking radiation has its roots in the properties of the
vacuum of a quantum field. Quantum fluctuations of the vacuum can create a
pair of particle-antiparticle straddling the black hole horizon. The
particles inside the horizon are confined in the black hole, while
outside particle can flow away to infinity.  Particle creation will produce states of the type
\be\label{state}\ket{\psi}_i=Ce^{\gamma\hat{b}^\dagger_i\hat{c}^\dagger_i}\ket{0}_{b_i}\ket{0}_{c_i}\ee
where $C$ is a normalization constant and $\gamma < 1$. Here, the $b_i$ modes live
inside the horizon, while the $c_i$ modes are outside the horizon.
The total state is given by
$\ket{\psi}=\otimes_{i=1}^N\ket{\psi}_i$.
We can write the Hilbert space for the quantum field as \be
\label{bhspace} \mathcal H = \mathcal H_{in}\otimes \mathcal
H_{out}. \ee It is easy to see that the total state $\ket{\psi}$ is
entangled in the bipartition of this space. The evolution for the partial state $\rho_{out}(t)$, outside the black hole
will not be unitary. When at time $t_f$ the whole black hole has completely evaporated, the full system will
be described by the density matrix $\rho_{out}(t_f)$, representing a
mixed state.
Here we see that starting with a pure state we end up with a mixed one even if the evolution is unitary, thus leading to a contradiction.

Let us now assume, as a possible way out, the existence of a quantum process that
completely disentangles the state $\ket{\psi}$, with respect to the natural bipartition before the black hole has completely evaporated.  We suggest here that the deformed structure of the Hilbert space described in (\ref{tensprod})
renders such a process viable. Let us consider the creation of $n$ pairs
with distribution in momenta $\vec{n}_R$ at a certain stage of black hole evaporation. Their Hilbert space can be written as
\be
\label{bhn}\mathcal H = \mathcal H_{in}\otimes (\mathcal S_n\mathcal
H^n \otimes \mathbb{C}^{N(\vec{n}_R)})_{out}\equiv \mathcal
H_{A}\otimes\mathcal H_{B}\otimes\mathcal H_{C}
 \ee
The dynamics of the process can be then seen as induced by an Hamiltonian of the form
\be\label{h}
H = H_{AB}+H_{AC}+H_{BC} .
\ee
The first term $H_{AB}$ describes the process of particle
creation. The other two terms are new. Now, let us assume that the three terms in Eq.(\ref{h}) act at different times: in the intervals $(t_0,t_1), (t_1,t_2)$ and $ (t_2,t_f)$ the process is dominated by  $H_{AB}, H_{AC}$ and $ H_{BC}$, respectively. Under this hypothesis, if we start from an initial state $\ket{0}$ , which is separable with respect to all the bipartitions in $\mathcal H$, we will have the final state
\be\label{unitary}
\ket{\psi (t_f)}\simeq
U_{BC}(t_f,t_2)\cdot U_{AC}(t_2,t_1)\cdot U_{AB}(t_1,t_0) \ket{0} .
\ee
At the beginning, the process is dominated by the first unitary $U_{AB}$ which creates an entangled pair with
respect to the bipartition $(A,B)$. Later, the second unitary $U_{AC}$ can completely disentangle
the state with respect to $(A,BC)$, {\it i.e.}, the separation {\em in-out},
and entangle it with respect to the bipartition $(B,C)$.
All the entanglement between the parties $A,B$ has been transferred to the parties $B,C$. Finally, the last unitary $U_{BC}$ will make the state evolve just outside of the event horizon. In this way we will have first
created an entangled state $\ket{\psi}_{AB}\otimes\ket{\psi}_C$ and
then we will have completely transferred the entanglement to obtain
a state $\ket{\psi_f}= \ket{\psi}_{A}\otimes\ket{\psi}_{BC}$.

In order for this to happen, the logarithm of the dimensions of the
Hilbert spaces $\mathcal H_{B},\mathcal H_{C}$ has to be greater than or
equal to the amount of entanglement initially created in the state
$\ket{\psi}_{AB}$.  A reasonable estimate of this entanglement can be
given by the following argument. Expanding (\ref{state})
we can safely estimate the amount of entanglement for any emitted pair with a
number of order unity, since higher terms in the expansion
will be negligible. The total entanglement of the state will thus be
given by an entropy $S \propto n$, where $n$ is the number of emitted quanta \cite{mathur}.
On the other hand, the logarithm of the dimension of the Hilbert
space $H_{C}$ grows faster than $n$. For instance, in the sector $n_i = 1\,\, \forall i$
it will be given by $\log_2 N \sim n\log_2 n$
and therefore all the entanglement can be in principle transferred.
Now we can wrap up the result. We start with a completely separable pure state. The process of evaporation entangles the state in
many ways, but the final state is a product respect to the
bipartition $A(B,C)$. The disappearance of the degrees of
freedom in $A$ does not change the purity of the state. Nevertheless the
state is seen as a thermal state by an observer that is not able to
resolve the Planck scale, because she is just observing a partial system.
The whole process is unitary.

{\em Conclusions and outlook.---} We have discussed how the introduction of non-local effects motivated by dynamical gravity in QFT can endow the Hilbert space of the theory with a richer structure. This structure renders possible entanglement between the modes of the field and allows in principle to ``hide" the unitarity of a quantum process in the fine structure of the new space. As an application, we have shown that in the new framework entanglement can be transferred away from the bipartition in-out for a radiating black hole.  The presence of this entanglement lies at the basis for the information paradox. We also want to remark that the additional Hilbert space $\mathcal H_C$ outside the black hole could also be used to ``lock" the information escaped from the black hole instead of hypothesizing a small remnant as proposed in \cite{lock}. In this letter, we have not provided a physical mechanism for the entanglement transfer, but we merely shown that there could be ``room" for it to happen. We hope that this might open a new avenue for future efforts toward a resolution of the black hole information paradox.

{\em Acknowledgments.---} We would like to thank Samuel Braunstein, Olaf Dreyer, Seth Lloyd, Valter Moretti, and Paolo Zanardi, for valuable discussion. Research at Perimeter Institute for Theoretical Physics is supported in part by the Government of Canada through NSERC and by the Province of Ontario through MRI. This project was partially supported by a grant from the Foundational Questions Institute (fqxi.org), a grant from xQIT at MIT. We also acknowledge financial support from DTO-ARO, ORDCF, CFI, CIFAR, and MITACS.

%:biblio


\begin{thebibliography}{99}

%\cite{Hawking:1974sw}
\bibitem{Hawking:1974sw}
  S.~W.~Hawking,
  %``Particle Creation By Black Holes,''
  Commun.\ Math.\ Phys.\  {\bf 43}, 199 (1975)
  [Erratum-ibid.\  {\bf 46}, 206 (1976)].

%\cite{Hawking:1976ra}
\bibitem{Hawking:1976ra}
  S.~W.~Hawking,
  %``Breakdown Of Predictability In Gravitational Collapse,''
  Phys.\ Rev.\  D {\bf 14}, 2460 (1976).


%\cite{Giddings:2006sj}
\bibitem{Giddings:2006sj}
  S.~B.~Giddings,
  %``Black hole information, unitarity, and nonlocality,''
  Phys.\ Rev.\  D {\bf 74}, 106005 (2006)
  [arXiv:hep-th/0605196].

%\cite{Rovelli:1990ph}
\bibitem{Rovelli:1990ph}
  C.~Rovelli,
  %``WHAT IS OBSERVABLE IN CLASSICAL AND QUANTUM GRAVITY?,''
  Class.\ Quant.\ Grav.\  {\bf 8}, 297 (1991).

%\cite{Dittrich:2005kc}
\bibitem{Dittrich:2005kc}
  B.~Dittrich,
  %``Partial and Complete Observables for Canonical General Relativity,''
  Class.\ Quant.\ Grav.\  {\bf 23}, 6155 (2006)
  [arXiv:gr-qc/0507106].

%\cite{Giddings:2005id}
\bibitem{Giddings:2005id}
  S.~B.~Giddings, D.~Marolf and J.~B.~Hartle,
  %``Observables in effective gravity,''
  Phys.\ Rev.\  D {\bf 74}, 064018 (2006)
  [arXiv:hep-th/0512200].

%\cite{ArkaniHamed:2007ky}
\bibitem{ArkaniHamed:2007ky}
  N.~Arkani-Hamed, S.~Dubovsky, A.~Nicolis, E.~Trincherini and G.~Villadoro,
  %``A Measure of de Sitter Entropy and Eternal Inflation,''
  JHEP {\bf 0705}, 055 (2007)
  [arXiv:0704.1814 [hep-th]].

%\cite{Bernard:1990ys}
\bibitem{Bernard:1990ys}
  D.~Bernard and A.~Leclair,
  %``Quantum group symmetries and nonlocal currents in 2-D QFT,''
  Commun.\ Math.\ Phys.\  {\bf 142}, 99 (1991).

%\cite{Arzano:2007nx}
\bibitem{Arzano:2007nx}
  M.~Arzano,
  %``Quantum fields, non-locality and quantum group symmetries,''
  Phys.\ Rev.\  D {\bf 77}, 025013 (2008)
  [arXiv:0710.1083 [hep-th]].

%\cite{AmelinoCamelia:2001fd}
\bibitem{AmelinoCamelia:2001fd}
  G.~Amelino-Camelia and M.~Arzano,
  %``Coproduct and star product in field theories on Lie-algebra
  %non-commutative space-times,''
  Phys.\ Rev.\  D {\bf 65}, 084044 (2002)
  [arXiv:hep-th/0105120].

%\cite{Chaichian:2004za}
\bibitem{Chaichian:2004za}
  M.~Chaichian, P.~P.~Kulish, K.~Nishijima and A.~Tureanu,
  %``On a Lorentz-invariant interpretation of noncommutative space-time and  its
  %implications on noncommutative QFT,''
  Phys.\ Lett.\  B {\bf 604}, 98 (2004)
  [arXiv:hep-th/0408069].

 %\cite{DiFrancesco:1997nk}
\bibitem{DiFrancesco:1997nk}
  P.~Di Francesco, P.~Mathieu and D.~Senechal,
  ``Conformal Field Theory,''
%\href{http://www.slac.stanford.edu/spires/find/hep/www?irn=3810356}{SPIRES entry}
{\it  New York, USA: Springer (1997) 890 p}

%\cite{Majid:1994cy}
\bibitem{Majid:1994cy}
  S.~Majid and H.~Ruegg,
  %``Bicrossproduct Structure Of Kappa Poincare Group And Noncommutative
  %Geometry,''
  Phys.\ Lett.\  B {\bf 334}, 348 (1994)
  [arXiv:hep-th/9405107].

%\cite{Arzano:2007ef}
\bibitem{Arzano:2007ef}
  M.~Arzano and A.~Marciano,
  %``Fock space, quantum fields and kappa-Poincar\'e symmetries,''
  Phys.\ Rev.\  D {\bf 76}, 125005 (2007)
  [arXiv:0707.1329 [hep-th]].

%\cite{Mathur:2008wi}
\bibitem{mathur}
  S.~D.~Mathur,
  %``What Exactly is the Information Paradox?,''
  arXiv:0803.2030 [hep-th].

%\bibitem{qibook} M. A. Nielsen and I. L. Chuang, Quantum Computation and Quantum Information, CUP, 2000.

\bibitem{lock} J.A.~Smolin and J.~Oppenheim, Phys. Rev. Lett. {\bf 96}, 081302 (2006)
%\bibitem{garay} Luis J. Garay, Phys. Rev. Lett. {\bf 80}, 2508 - 2511 (1998)
%\bibitem{swap} M.~Zukowski, A.~Zeilinger, M.A.~Horne, and A.K.~Eckert, Phys. Rev. Lett. {\bf 71}, 4287 (1993)


\end{thebibliography}
\end{document}